\documentclass[a4paper,twoside]{article}
\usepackage{cite}
\usepackage{amsmath,amssymb,amsfonts}
\usepackage{algorithmic}
\usepackage{graphicx}
\usepackage{textcomp}
\usepackage{xcolor}
\usepackage{booktabs}

\usepackage{multirow}
\usepackage{url}
\usepackage{cite}
\usepackage[latin1]{inputenc}

\usepackage{glossaries}

\newacronym{OFDMA}{OFDMA}{Orthogonal Frequency Division Multiple Access}
\newacronym{SC-FDMA}{SC-FDMA}{Single Carrier- Frequency Division Multiple Access}
\newacronym{DCI}{DCI}{Downlink Control Information}
\newacronym{LSTM}{LSTM}{Long-Short Term Memory}
\newacronym{RMSE}{RMSE}{ Root Mean Squared Error}
\newacronym{MAE}{MAE}{ Mean Absolute Error}
\newacronym{RAN}{RAN}{Radio Access Network}
\newacronym{BS}{BS}{Base Station}
\newacronym{MNO}{MNO}{Mobile Network Operators}
\newacronym{M2M}{M2M}{Machine-to-Machine}
\newacronym{PDCCH}{PDCCH}{Physical Downlink Control Channel}
\newacronym{AI}{AI}{Artificial Intelligence}
\newacronym{RNN}{RNN}{Reccurent Neural Network}
\newacronym{MCS}{MCS}{Modulation and Coding Scheme}
\newacronym{PRB}{PRB}{Physical Resource Block}
\newacronym{RB}{RB}{Resource Block}
\newacronym{TBS}{TBS}{Transport Block Size}
\newacronym{RNTI}{RNTI}{Radio Network Temporary Identifier}
\newacronym{QoS}{QoS}{Quality of Service}
\newacronym{CRC}{CRC}{Cyclic Redundancy Check}
\newacronym{UE}{UE}{User Equipment}
\newacronym{TTI}{TTI}{Transmission Time Interval}
\newacronym{SF}{SF}{sub-frame}
\newacronym{ANN}{ANN}{Artificial Neural Network}
\newacronym{eNB}{eNB}{eNodeB}

\def\BibTeX{{\rm B\kern-.05em{\sc i\kern-.025em b}\kern-.08em
    T\kern-.1667em\lower.7ex\hbox{E}\kern-.125emX}}

\begin{document}

\title{Traffic Load Prediction and Power Consumption Reduction for Multi-band Networks \\
\thanks{This work has been partially supported by grant ANR-21-CE25-0005 from the Agence Nationale de la Recherche, France for the SAFE project.}
}

\author{
\footnotesize
Ndolane Diouf
\textit{Cheikh Anta Diop University}
Dakar, Senegal
ndolane.diouf@ucad.edu.sn
\and
\footnotesize
Cesar Vargas Anamuro
\textit{IMT Atlantique, IRISA}
Rennes, France
cesar.vargasanamuro@imt-atlantique.fr
\and
\footnotesize
Cedric Gueguen
\textit{University of Rennes 1, IRISA}
Rennes, France
cedric.gueguen@irisa.fr
\and
\footnotesize
Massa Ndong
\textit{Senegal Virtual University}
Dakar, Senegal
massandong@gmail.com
\and
\footnotesize
Kharouna Talla
\textit{Cheikh Anta Diop University}
Dakar, Senegal
kharouna.talla@ucad.edu.sn
\and
\footnotesize
Xavier Lagrange
\textit{IMT Atlantique, IRISA}
Rennes, France
xavier.lagrange@imt-atlantique.fr
}

\maketitle

\begin{abstract}
Energy is a major expense issue for mobile operators. In the case of wireless networks, base stations have been identified as the main source of energy consumption. In this paper, we study the energy consumption reduction problem based on real measurements for a commercial multi-band LTE network. Specifically, we are interested in sleep modes to turn off certain frequency bands during low traffic periods and consequently reduce power consumption. We determine the number of frequency bands really needed at each time period. The frequency bands that are not needed can be disabled to reduce energy consumption. In order to allow the operator to predict how many bands can be switched off  without major impact on the quality of service, we propose to use a deep learning algorithm, such as Long-Short Term Memory (LSTM). Based on the captured data traces, we have shown that the proposed LSTM model can save an average of 8\% to 21\% of the energy consumption during working days.
\end{abstract}

\section{Introduction}
The rise in demand and the emerging needs of mobile users have pushed \gls{MNO} to employ various techniques to increase the capacity of their cellular networks. One of these techniques involves the utilization of multiple frequency bands within the same cellular sector.

In mobile networks, \gls{BS} have been identified as the main source of energy consumption in \gls{RAN}, approximately 60\% to 80\% of the total energy consumption in cellular networks comes from \gls{BS}s \cite{kyuho2012toward}, \cite{marsan2009optimal}. Power consumption is even higher when the \gls{MNO} use multi-band. The introduction of sleep modes to turn off the \gls{BS}s during low traffic periods and turn them back on when traffic increases, has been identified as the most effective approach to save energy \cite{donevski2019neural}. To address this approach, it is important for the \gls{MNO} to be aware of traffic demands. Traffic analysis to meet user demand is important for the development of an intelligent mobile network. Today, the use of \gls{AI}, especially deep learning is interesting because it adapts to any situation in a mobile network and will allow the \gls{MNO} to get rid of the configuration. Deep learning has the ability to learn features at multiple levels of abstraction and allows a system to learn complex functions and maps the input into the output directly from the data \cite{wang2017deep}.

Several research works have proposed different energy reduction strategies for mobile networks. The authors of \cite{5683654} and \cite{6489498} both explore dynamic \gls{BS} switching to reduce energy consumption, with a focus on time-varying traffic characteristics and energy minimization problem formulation, respectively. In \cite{6335345}, the authors propose a solution to a maximization problem for turning off \gls{BS}s in mobile networks. Their scheme exploits the existence of idle periods when the traffic load is low. Their work focuses on finding the optimal combination of switched-on and switched-off \gls{BS}s leading to maximum energy savings. The authors in \cite{murthy2012survey} study the energy consumption by a typical \gls{BS} and attempt to examine possible energy-efficient solutions towards a green \gls{BS}. The authors of \cite{donevski2019neural} exploit the performance of \gls{ANN} for traffic prediction and for selecting the times to turn on and off small \gls{BS}s. First, they estimate the short-term traffic load, and from this derive the best combination of switching decisions. Then, they perform both traffic estimation and switching optimization. They focus on a heterogeneous network scenario with single-frequency micro and macro BSs. In contrast, our study centers around a scenario where the MNO uses multiple frequency bands within the same BS. The authors of \cite{trinh2018mobile} study the mobile traffic of a \gls{BS} and design a traffic prediction system using \gls{LSTM}. Similarly, our study utilizes information from the PDCCH. However, in addition to this, we aim to estimate the number of frequency bands required to reduce base station energy consumption. To the best of our knowledge, no previous studies have considered the multi-band characteristic of all operational networks, which opens the way for simple but efficient energy reduction methods.

To reduce energy consumption, we first propose an algorithm that determines the required number of frequency bands based on the traffic load. With prior knowledge of the number bands needed for network usage,  the \gls{MNO} can turn off unnecessary bands and thus reduce energy consumption. Additionally, we utilize \gls{LSTM} model to develop a predictive algorithm for determining the number bands required in the LTE network. To the best of our knowledge, we are the first to consider a multi-band network to propose an energy reduction algorithm and a deep learning algorithm to enable/disable a frequency band.

The rest of the article is organized like this. Section~\ref{Data_Set} provides details on the LTE control channel and datasets. The proposed algorithm for energy reduction is given in Section~\ref{Proposed_algorithm}. Section~\ref{LSTM_Model} contains a detailed explanation of the deep learning approach that is tested as a solution to the problem. The potential energy consumption reduction is detailed in Section~\ref{Energy_consumption_reduction}. Finally, Section~\ref{Conclusion} concludes the article.

\section{User Traffic Data Set of a Commercial Multiband LTE System}
\label{Data_Set}

\subsection{LTE Control Channel}

The allocation of resources in LTE is done in both the time and frequency domains. Time is divided into frames of 10~ms each. In the same way, a radio frame is divided into 10 \glspl{SF} called \gls{TTI} of 1~ms duration each. In the frequency domain, the bandwidth is divided into sub-bands of 180~kHz. Each of these sub-bands is composed of 12 sub-carriers regularly spaced of 15~kHz. The physical resources in the time and frequency domain are called \gls{RB} \cite{ezzaouia:tel-02118466}.

The eNodeB uses an \gls{RNTI} to identify terminals inside the cell. The \gls{DCI} is transmitted via the \gls{PDCCH}, which is sent at the beginning of each downlink \gls{SF}. The \gls{DCI} contains the \gls{MCS}, the number of  \gls{PRB} and the \gls{TBS} assigned to the terminals every millisecond. The \gls{RNTI} of the addressed terminal scrambles the \gls{CRC} of the \gls{DCI}.

\subsection{Datasets}
\label{dataset}

The collected data set consists of two weeks of observation, from February 19, 2022 to March 04, 2022 that we collected by monitoring four frequency bands (i.e., 800~MHz, 1800~MHz, 2100~MHz, and 2600~MHz) of an LTE cell of the \gls{MNO} SFR located in the city of Rennes, France. The dataset used for this study comprises all \gls{DCI} control messages transmitted by the \gls{BS}.

\section{Proposed Algorithm for Energy Reduction}
\label{Proposed_algorithm}

The proposed algorithm attempts to determine the number of frequency bands needed based on the network load. Frequency bands that are not needed can be disabled to reduce energy consumption. The main idea is to transfer users connected in the high frequency bands to the low frequency bands when the network usage is low. The reason for this is that the coverage is better in the low frequencies than in the high frequencies. Moreover, the algorithm takes into account the \gls{QoS} of users. Our objective is that the additional delay induced by switching off some bands is limited to a maximum value, we consider 20~ms in this study.

\begin{figure}[!t]
  \centering
   \includegraphics[width=0.6\textwidth]{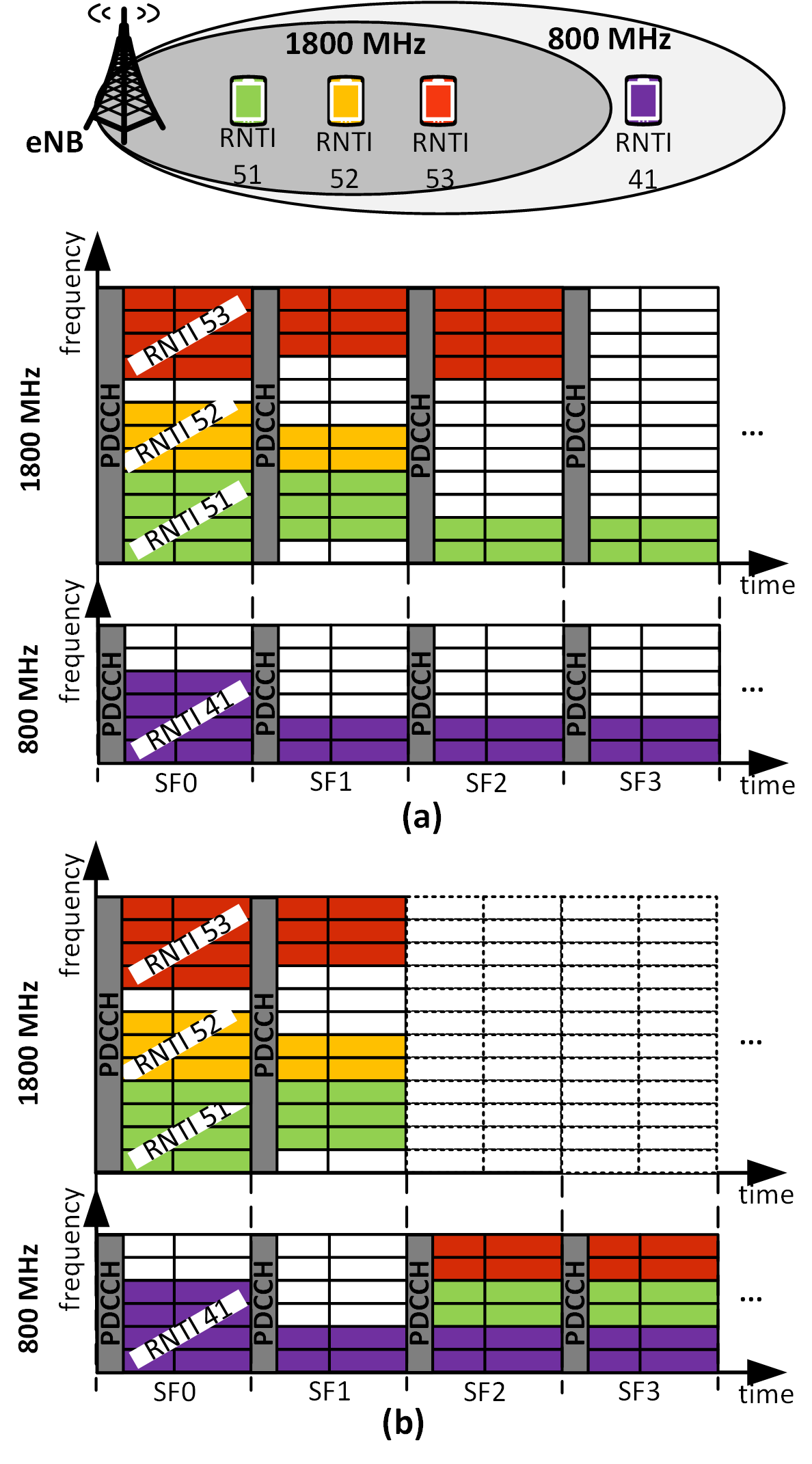}
    \caption{Resource allocation for two frequency bands: (a) traditional and (b) reallocation using the proposed algorithm.}
    \label{fig:PRB_transfer}
\end{figure}

Fig.~\ref{fig:PRB_transfer} shows a simplified example of the transfer of users between frequency bands. In this example, we consider two bands: 800~MHz and 1800~MHz, which have 6 and 12 PRBs per sub-frame, respectively. We only display the allocation of resources in the downlink since it is typically more utilized than the uplink. Fig.~\ref{fig:PRB_transfer}~(a) shows the resource allocation considering that both frequency bands are active. Three terminals are camping on the 1800~MHz band, and one terminal is camping on the 800~MHz band. Some PRBs are unallocated to any terminal (white PRBs). Fig.~\ref{fig:PRB_transfer}~(b) shows the resource reallocation using the proposed algorithm. We define the reallocation period as the period during which the scheduling and allocation can be modified. In this example, we consider a reallocation period of 2~ms, and therefore, the 800~MHz band has a maximum capacity of 12 PRBs per reallocation period. In the first reallocation period (SF0 and SF1), a total of 24 PRBs are used, requiring both frequency bands. Only 12 PRBs are used in the next reallocation period (SF2 and SF3), therefore, only the 800~MHz band is required. The resources that were previously allocated to the 1800~MHz band are now reallocated to the 800~MHz band. It should be noted that two PRBs allocated to RNTI-53 (red) are delayed by one millisecond.

Let $F$ be the number of frequency bands available in the cell and $f \in \{1,2,...,F\}$ denotes the frequency band number, where $f=1$ is the lowest frequency band and $f=F$ is the highest frequency band. Our data set includes for each band and each \gls{SF} (1-ms granularity) how many \glspl{PRB} are allocated. The first step of the study is to count how many bands were really necessary when a moderate additional delay is tolerated. The activation period is the minimum time interval during which the frequency band will remain on or off. The duration of the activation period $T$ (e.g. 10 minutes) will depend on the \gls{MNO} ability to switch the frequency band on/off and off/on. For each activation period $t$, we determine the number of required frequency bands $N_t$ by following these steps:
\begin{enumerate}
    \item Each activation period $t$ is divided into reallocation periods of a few milliseconds (e.g. 20~ms). Let $I$ be the number of reallocation periods in an activation period. So $I=T/20$, where $T$ is the duration of the activation period in ms.
    \item For each reallocation period $i$ the algorithm determines the number of required bands. To do so, it compares the total number of allocated PRBs\footnote{The total number of allocated PRBs includes all bands in the cell.} and the number of available PRBs in the cell. The number of required frequency bands $n_{t,i}$ in the reallocation period $i$ of the activation period $t$ is given by:
    \begin{equation}
      n_{t,i} =
        \begin{cases}
          1 & \text{if $\theta_{t,i} \leq S_1$}\\
          2 & \text{if $S_1 < \theta_{t,i} \leq S_2$}\\
          3 & \text{if $S_2 < \theta_{t,i} \leq S_3$}\\
          4 & \text{otherwise}
        \end{cases}
        \label{eq_1}
    \end{equation}
    where $\theta_{t,i}$ is the total number of PRBs allocated in the reallocation period $i$ of the activation period $t$; $S_1$, $S_2$, $S_3$ are the activation thresholds for one, two and three frequency bands, respectively. These activation thresholds depend on the number of available PRBs per band. $S_j = \delta\sum\limits_{f = 1}^{j} A_f$, where $\delta$ is the duration of the reallocation period in milliseconds and $A_f$ is the number of PRBs per TTI in the band $f \in \{1, 2, 3, 4\}$.

    \item The number of frequency bands required in the activation period $t$ is $N_t =\max \{n_{t,0}, n_{t,1}, ..., n_{t,I-1}\}$.

\end{enumerate}

\section{A Long-Short Term Memory Network}
\label{LSTM_Model}

\subsection{LSTM Architecture}
LSTM-based models extend the memory of \gls{RNN}s to allow them to retain and learn long-term dependencies of inputs and solve the leakage gradient problem \cite{gers2000learning}. Their learning ability is due to the structure of the \gls{LSTM} units each consisting of a forget gate, an input gate, and an output gate. In addition, the cell state represents the memory of the unit.

\begin{figure}[b]
 \centering
 \includegraphics[scale=0.5]{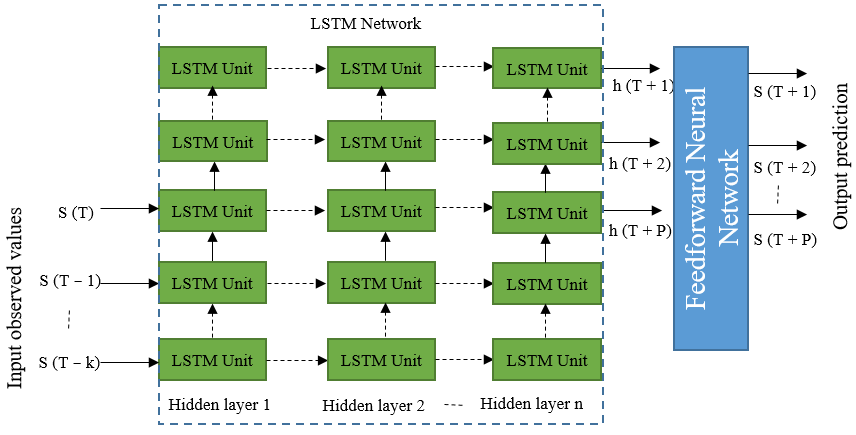}
    \caption{Proposed architecture for frequency band number prediction}
    \label{fig:LSTM_Network}
\end{figure}

Fig.~\ref{fig:LSTM_Network} illustrates the proposed architecture for the prediction of the number of frequency bands. This architecture is a stacked \gls{LSTM} network obtained by grouping several layers of basic \gls{LSTM} units. The \gls{LSTM} unit of each layer extracts a fixed number of features that are passed to the next layer. In our architecture, the input is the vector $s$, which contains a set of required frequency band numbers. The number of observations is the number of selected time intervals $T_x$ which is the number of \gls{TTI} for which the traffic is aggregated. In the expression for $s(T-k)$, $k$ represents a fixed number of past values to predict the number of bands for the next time slot.

The principle used by our \gls{LSTM} network for prediction is to observe a number of frequency bands during a fixed number of time intervals up to $T$, and then tries to predict the number of frequency bands in the next time interval $T + 1$.  Then, the output of the \gls{LSTM} network is fed to a dense layer of fully connected neuron, which gives the final prediction of number of frequency bands.

\subsection{Predictive Methodology}

We use the mobile traffic dataset collected on a \gls{BS} (refer to Section~\ref{dataset}) to evaluate the performance of our proposed architecture. Granularities of 1~minute, 3~minutes, 10~minutes, 30~minutes, and 1~hour are used to predict the number of frequency bands really needed at each activation period. We implement the algorithm in python, using Keras and Tensorflow as backend. The hyperparameters reported in Table~\ref{table_3} are chosen taking into account the above granularities. In this table, the initial learning rate hyperparameter controls how quickly the model fits the problem. Batch size refers to the number of training examples used in an iteration. After testing a variable number of hidden layers, we set this number to 6 which gives us more precision. An epoch indicates the number of runs of the full training dataset that the machine learning algorithm has performed. Adam's optimizer \cite{kingma2014adam} is used to iteratively update the network weights based on the training data. The dataset used for this study spans a period of two weeks. For the prediction of the number of required frequency bands, the data from the five working days of the first week are used to train and validate the LSTM model. We do not consider weekends due to insufficient data on these days. Therefore, the results presented in this study are optimistic, as weekends typically experience low traffic and thus significant energy savings can be achieved. The forecasts are linked with the first two working days of the second week.

\begin{table}[]
\renewcommand{\arraystretch}{1.0}
\setlength{\tabcolsep}{0.09 cm}
\centering
\caption{Training Hyperparameters}
    \begin{tabular}{c cccccc}
    \toprule &
    \multicolumn{5}{c}{\textbf{Activation Period}} \\
    \cmidrule[0.4pt](lr{0.125em}){2-6}
    Hyperparameters & 1 min & 3 min & 10 min & 30 min & 1 h \\
    \midrule
    Initial learning rate & 0.0001 & 0.0001 & 0.0001 & 0.0001 & 0.0001 \\
    \hline Number of epochs & 100 & 100 & 100 & 100 &  150 \\
    \hline LSTM hidden states & 256 & 256 & 256 & 256 & 256 \\
    \hline LSTM hidden layers  & 6 & 6 & 6 & 6 & 6 \\
    \hline Feedforward hidden layer & 1 & 1 & 1 & 1 & 1 \\
    \hline Batch size & 72 & 16 & 16 & 2 & 2 \\
    \hline Optimization algorithme & Adam & Adam & Adam & Adam & Adam \\
    \hline Loss Function & RMSE & RMSE & RMSE & RMSE & RMSE \\
    \hline
    \label{table_3}
    \end{tabular}
\end{table}

\subsection{Evaluation Metrics}

We use \gls{RMSE}, accuracy, and QoS preservation as metrics to measure the performance of our proposed architecture.

The \gls{RMSE} represents the error of the predictive model. It is used to measure the quality of the model based on the predictions made on the training dataset compared to the true label. The lower the value, the better the model. The \gls{RMSE} is defined by

\begin{equation}
RMSE = \sqrt{\sum\limits_{t = 1}^{N}(\tilde{x}_t - x_t)^2/N}
\end{equation}
where $N$ is the total number of points, $\tilde{x}_t$ and $x_t$ are respectively the predicted value and the actual values at time $t$.

Our model predicts either 1, 2, 3, or 4 bands, which allows us to calculate the accuracy. The accuracy is defined by

\begin{equation}
accuracy = \sum\limits_{t = 1}^{N}f(x_t, \tilde{x}_t)/N
\end{equation}
where $f(x_t, \tilde{x}_t)=1$ if $\tilde{x}_t$ = $x_t$ and 0 otherwise.

Two scenarios are possible when forecasting the number of required frequency bands:
\begin{enumerate}
    \item If $\tilde{x}_t \geq x_t$, our model is overestimating the required number of bands. In this case, the model does not optimize the energy consumption but allows the \gls{MNO} to ensure the user \gls{QoS}.
    \item If $\tilde{x}_t < x_t$, our model predicts less than the required number of bands. Thus, our model optimizes energy consumption but does not ensure \gls{QoS}. The \gls{MNO} will delay the packets until there are \gls{PRB}s available. This time delay could be greater than 20~ms, which can impact the user \gls{QoS}.
\end{enumerate}

Since predicting more frequency bands than necessary does not impact \gls{QoS}, we propose a new metric called \gls{QoS} preservation, obtained by modifying the accuracy formula. The QoS preservation is defined as the percentage of cases where the predicted number of bands is greater than or equal to the required number.

\begin{equation}
QoS~preservation = \sum\limits_{t = 1}^{N}
{f}^{*}(x_t, \tilde{x}_t)/N,
\end{equation}
where $f^{*}(x_t, \tilde{x}_t)=1$ if  ${x}_t \leq \tilde{x}_t$ and 0 otherwise.

\subsection {Forecasting the Number of Frequency Bands}

\begin{figure}[h]
    \centering
    \includegraphics[scale=0.45]{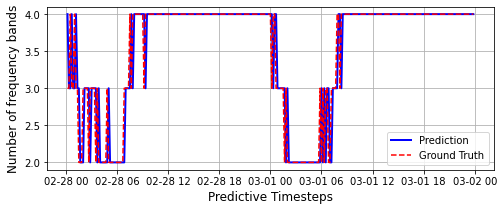}
    \caption{Number of frequency band forecasted vs. ground-truth measurements for two days (activation period = 10 minutes).}
    \label{fig:Forcasted_10min}
\end{figure}

After implementing the \gls{LSTM} model, we use a training and test or validation dataset to measure the performance of our model. The \gls{RMSE} is used to examine training and validation losses. As an example, in Fig.~\ref{fig:Forcasted_10min}, we show the prediction of the number of frequency bands needed for a period of two days (February 28th and March 1st) considering an activation period of 10 minutes. We can see that the prediction is fairly accurate compared to the ground truth.

The performance of our model for predicting the number of required bands for each activation period is reported in Table~\ref{performance_lstm}. The results show that the LSTM model performance improves as the activation period increases, reaching a maximum with an activation period of 10~minutes. However, after that, performance began to decrease.

\begin{table}[]
    \renewcommand{\arraystretch}{0.7}
    \caption{Performance of the \gls{LSTM} model.}
    \centering
    \begin{tabular}{c ccccc}
    \toprule &
      \multicolumn{5}{c}{\textbf{Activation Period}} \\
      \cmidrule[0.4pt](lr{0.125em}){2-6}
    Metric & 1 min  & 3 min & 10 min & 30 min & 1 h \\
    \midrule
    RMSE  & 0.534 & 0.413 & 0.354 & 0.383 & 0.412 \\
    Accuracy (\%) & 71.58 & 82.89 & 87.45 & 88.42 & 82.97  \\
    QoS~preservation (\%) & 85.75 & 91.55 & 93.72 & 93.68 & 91.48 \\
    \label{performance_lstm}
    \end{tabular}
\end{table}

\section{Reduction of Energy Consumption}
\label{Energy_consumption_reduction}

In this section, we present the statistical analysis results obtained from the algorithm proposed in Section~\ref{Proposed_algorithm} and the predictive analysis results obtained from our \gls{LSTM} model proposed in Section~\ref{LSTM_Model}. The percentage of energy savings is directly related to the period of time that the frequency bands are in sleep mode. Furthermore, we estimate the average extra delay\footnote{The average extra delay is the total delay of the reallocated PRBs divided by the total number of allocated PRBs in the cell.} caused by the proposed algorithm and the LSTM model.

\subsection{Reference Strategy Results}
Using the proposed algorithm, we analyze the maximum period of time that the frequency bands can be in sleep mode, assuming perfect knowledge of the incoming traffic.
In Table~\ref{week1_results_ideal}, we present the percentage of time in sleep mode for each band during a given activation period for the first week (only working days). We note that a longer activation period results in a lower percentage of sleep time (i.e., lower energy savings) and a lower average extra delay. To explain these results: Recall that our algorithm divides each activation period into 20-ms reallocation periods. The number of required bands during an activation period is equal to the maximum number of required bands during any of the constituent reallocation periods. As an example, we analyze the number of required bands for one hour, during which there is a peak use of four bands and the rest of the time only one band is needed. For an activation period of 1 hour: during 60~minutes, four bands will be used while for an activation period of 1~minute: four bands will be needed during 1 minute and one band during 59 minutes. So, it is obvious that using an activation period of 1 minute would save much more energy than using a period of 1 hour. Hence, smaller activation periods result in a more accurate estimate of the number of required bands, while larger activation periods tend to overestimate the number of required bands. The average delay decreases with longer activation periods because overestimation results in fewer delayed PRBs compared to the reference model. If the LTSM model predicts four frequency bands during activation, none of the allocated PRBs will be delayed in that period. The worst-case scenario shows an average extra delay of only 159 microseconds.

Note that activation periods of 20~ms and 1~s are technically impossible to use in practice. These activation periods are presented in this study only as an indication. On the one hand, because the System Information Block (SIB) messages are sent every 80~ms \cite{9013397} and on the other hand because it takes time for the \gls{BS} to switch on and off. Henceforth, we will solely analyze activation periods greater than or equal to one minute.

\begin{table}[]
\renewcommand{\arraystretch}{0.6}
\setlength{\tabcolsep}{0.1 cm}
\centering
\caption{Reference sleep time percentage and average extra delay: first week data.}
\begin{tabular}{l ccccccc}
\toprule &
  \multicolumn{7}{c}{\textbf{Activation Period}} \\
  \cmidrule[0.4pt](lr{0.125em}){2-8}
  Band & 20 ms & 1 s & 1 min &  3 min &  10 min &  30 min &  1 h\\
\midrule
800 MHz (\%)  & 0 & 0 & 0 & 0 & 0 & 0 & 0 \\
1800 MHz (\%)  & 51.57	& 22.82	& 3.54	& 1.67	& 0.28	& 0	 & 0 \\
2100 MHz (\%) & 93.87	& 64.79	& 25.25	& 20.88	& 16.39	& 11.25 & 8.33 \\
2600 MHz (\%)  & 99.90	& 97.81	& 63.42	& 45.96	& 34.58 & 28.33 & 25 \\
\midrule
Average sleep (\%) & 61.33 & 46.36 & 23.05 & 17.13 & 12.81 & 9.90 & 8.33 \\
Average delay ($\mu$s) & 159.9 & 15.9 & 0.539 & 0.14 & 0.027 & 0.004 & 0.002 \\
\label{week1_results_ideal}
\end{tabular}
\end{table}

The percentage of sleep time for each frequency band for a given activation period for the first two working days of the second week is presented in Table~\ref{week2_results_ideal}. We note that the trends are similar to those obtained in the first week.

\begin{table}[]
 \renewcommand{\arraystretch}{0.6}
 \setlength{\tabcolsep}{0.1 cm}
 \centering
 \caption{Reference sleep time percentage and average extra delay: second-week data.}
\begin{tabular}{c ccccc}
 \toprule &
   \multicolumn{5}{c}{\textbf{Activation Period}} \\
   \cmidrule[0.4pt](lr{0.125em}){2-6}
Band & 1 min & 3 min & 10 min & 30 min & 1 h \\
 \midrule
 800 MHz (\%)  & 0 & 0 & 0 & 0 & 0 \\
1800 MHz (\%)  & 1.53 & 0.10 & 0 & 0 & 0  \\
 2100 MHz (\%)  & 25.45 & 21.77 & 17.36 & 12.5 & 8.33 \\
 2600 MHz (\%)  & 56.22  & 39.79 & 30.90 & 26.04 & 22.92 \\
 \midrule
 Average sleep (\%) & 20.80 & 15.42 & 12.07 & 9.64 & 7.81 \\
Average delay ($\mu$s) & 0.579 & 0.121 & 0.022 & 0.002 & 0.001  \\
 \label{week2_results_ideal}
 \end{tabular}
 \end{table}

\subsection{Predictive Analysis Results}

In this section, we present the percentage of time that each band spends in sleep mode. This is estimated by using the LSTM model's predictions for the first two working days of the second week. The results for different activation periods are presented in Table~\ref{week2_results_lstm}. Regarding the percentage of sleep time, the results are similar to those obtained with the proposed algorithm assuming perfect knowledge of the incoming traffic (reference strategy). Regarding the average extra delay, the trend is similar to the results obtained using the proposed algorithm; however, the values are higher. This is because when the LSTM model predicts a lower number of bands than required, there may be many more PRBs that need to be reallocated.

\begin{table}[]
\renewcommand{\arraystretch}{0.6}
\setlength{\tabcolsep}{0.1 cm}
\centering
\caption{LSTM model sleep time percentage and average delay: second-week data}
\begin{tabular}{c ccccc}
\toprule &
  \multicolumn{5}{c}{\textbf{Activation Period}} \\
  \cmidrule[0.4pt](lr{0.125em}){2-6}
Band & 1 min  & 3 min &  10 min & 30 min & 1 h   \\
\midrule
800 MHz (\%)    & 0 & 0 & 0 & 0 & 0 \\
1800 MHz (\%)  & 1.53 & 0 & 0 & 0 & 0 \\
2100 MHz (\%)  & 25.46  &  21.79 & 17.42 & 12.63 &  8.51 \\
2600 MHz (\%)  & 56.16  &  39.72 & 31.01 & 26.31  & 23.40  \\
\midrule
Average sleep (\%) & 20.78 & 15.37 & 12.10 & 9.73 & 7.97  \\
Average delay ($\mu$s) & 11.086 & 1.014 & 0.179 & 0.097 & 0.064   \\

\label{week2_results_lstm}
\end{tabular}
\end{table}

\subsection{Energy Saving}
We consider two energy consumption models, which are shown in Table~\ref{ec_models}. In model~1, all frequency bands consume the same energy, while in model~2, energy consumption is proportional to the bandwidth of the frequency band.

\begin{table}[]
\caption{Energy consumption models}
\centering
\begin{tabular}{c ccc}
\hline
Band & Bandwidth & Model 1 & Model 2 \\
\hline
800 MHz   & 10 MHz & P& P \\
1800 MHz  & 20 MHz & P& 2P \\
2100 MHz  & 15 MHz & P& 1.5P\\
2600 MHz  & 15 MHz & P& 1.5P\\
\label{ec_models}
\end{tabular}
\end{table}

The energy saving $\rho$ is given by
\begin{equation}
\rho = \sum\limits_{f = 1}^{4}\beta_f P_f/\sum\limits_{f = 1}^{4}P_f,
\label{eq_consommation}
\end{equation}

where $\beta_f$ and $P_f$ are respectively the sleep time percentage and the energy consumption with respect to the frequency band $f$. Recall that the 800~MHz band is always active, i.e., $\beta_1=0$. If the \gls{MNO} keeps all frequency bands active (i.e., $\beta_f = 0$ for all $f$), thus $\rho =0$.

\begin{figure}[]
    \centering
    \includegraphics[scale=0.5]{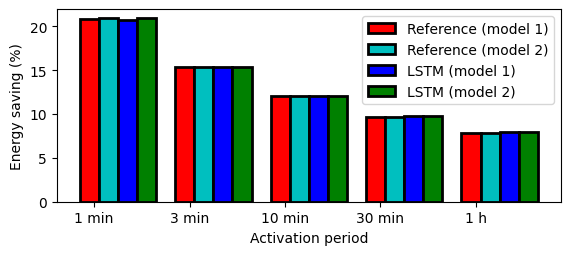}
    \caption{Reference strategy vs. energy saving with LSTM model.}
    \label{fig:Energy_saving}
\end{figure}

In Fig.~\ref{fig:Energy_saving}, we represent the energy saved using the reference strategy and our proposed LSTM model. The energy consumption of the \gls{BS} can be reduced by 8\% to 21\% depending on the activation period. Both energy consumption models yield very similar results. The reason is that the 1800~MHz band is almost always active (i.e., $\beta_2 \approx 0$). Substituting these values into Eq.~\ref{eq_consommation}, we obtain that the energy savings are the same for both models. Furthermore, the results indicate that as the activation period increases, the amount of energy savings decreases. These results are expected since energy savings are correlated with the percentage of sleep time. The LSTM model gives results very close to those obtained by the reference strategy. However, for activation periods greater than 1~minute, our LSTM model gives slightly better results than the reference. The reason is that our model predicts fewer bands than the reference and thus it saves more energy.

\begin{figure}[]
    \centering
    \includegraphics[scale=0.5]{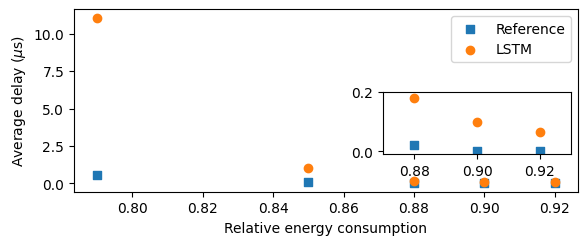}
    \caption{Relative energy consumption versus average extra delay.}
    \label{fig:Energy_vs_delay}
\end{figure}

Fig.~\ref{fig:Energy_vs_delay} shows  the relative energy consumption, defined as the ratio between energy consumption with an energy-saving model and energy consumption without it, versus the average extra delay. The LSTM model consistently exhibits higher delay than the reference strategy for all activation periods.This is due to the LSTM model sometimes predicting fewer frequency bands than required (see QoS preservation in Table~\ref{performance_lstm}), resulting in a higher extra delay. Conversely, there is a trade-off between extra delay and energy consumption in both cases. Reducing the delay leads to an increase in relative energy consumption in both cases.

\section{Conclusion}
\label{Conclusion}
In this paper, we study the energy consumption reduction problem for a commercial multi-band LTE network. Specifically, we are interested in sleep modes to turn off certain bands during low traffic periods and consequently reduce energy consumption, without affecting network coverage or user QoS. To achieve this, we proposed an algorithm and a \gls{LSTM} model. The proposed algorithm seeks to determine the number of required bands based on the traffic load. Bands that are not needed can be turned off to reduce energy consumption. The proposed LSTM model adapts to predict the load level to selectively turn on the bands to be used. The performance evaluation performed on a real data set of a multi-band network demonstrates the validity of the proposed \gls{LSTM} model, achieving a maximum accuracy of 88.42\%, and a QoS preservation of 93.72\%. Our LSTM model can save an average of 8\% to 21\% of energy consumption during working days. As future work, we plan to consider weekends, days in which the savings could be even greater due to low traffic.

\bibliographystyle{plain}
\bibliography{biblio}
\end{document}